\documentclass[aps,prl,a4paper,twocolumn,superscriptaddress,citeautoscript,footinbib]{revtex4-1}
\usepackage{color}
\usepackage[english]{babel}
\usepackage{lmodern}
\usepackage[T1]{fontenc}
\usepackage{amsmath}
\usepackage{units}
\usepackage{grffile}
\usepackage{color}
\usepackage[bottom]{footmisc}
\usepackage{booktabs}
\usepackage{dcolumn}
\usepackage{graphicx}
\usepackage{lipsum}
\graphicspath{{figs/}}

\newcommand\beq{\begin{equation}}
\newcommand\eeq{\end{equation}}

\newcommand{\J}{J}

\def\SRO{Sr$_2$RuO$_4$\,}
\newcommand{\TFL}{$T_{\mathrm{FL}}$\,}
\newcommand{\leff}{\lambda^{\mathrm{eff}}}
\newcommand{\abar}{\underline{a}}
\newcommand{\bbar}{\underline{b}}

\begin{document}
\title{Spin-Orbit Coupling and Electronic Correlations in Sr$_{2}$RuO$_{4}$}
\author{Minjae Kim}
\email{garix.minjae.kim@gmail.com}
\affiliation{Centre de Physique Th\'eorique, \'Ecole Polytechnique, CNRS, Universit\'e Paris-Saclay, 91128 Palaiseau, France}
\affiliation{Coll\`ege de France, 11 place Marcelin Berthelot, 75005 Paris, France}
\author{Jernej Mravlje}
\affiliation{Jo\v{z}ef Stefan Institute, Jamova 39, Ljubljana, Slovenia}
\author{Michel Ferrero}
\affiliation{Centre de Physique Th\'eorique, \'Ecole Polytechnique, CNRS, Universit\'e Paris-Saclay, 91128 Palaiseau, France}
\affiliation{Coll\`ege de France, 11 place Marcelin Berthelot, 75005 Paris, France}
\author{Olivier Parcollet}
\affiliation{Institut de Physique Th\'eorique (IPhT), CEA, CNRS, 91191 Gif-sur-Yvette, France}
\affiliation{Coll\`ege de France, 11 place Marcelin Berthelot, 75005 Paris, France}
\author{Antoine Georges}
\affiliation{Coll\`ege de France, 11 place Marcelin Berthelot, 75005 Paris, France}
\affiliation{Center for Computational Quantum Physics, Flatiron Institute, 162 Fifth avenue, New York, NY 10010, USA} 
\affiliation{Centre de Physique Th\'eorique, \'Ecole Polytechnique, CNRS, Universit\'e Paris-Saclay, 91128 Palaiseau, France}
\affiliation{Department of Quantum Matter Physics, University of Geneva, 24 Quai Ernest-Ansermet, 1211 Geneva 4, Switzerland}

\date{\today}
\begin{abstract}
We investigate the interplay of spin-orbit coupling (SOC) and
electronic correlations 
in Sr$_{2}$RuO$_{4}$ using dynamical mean-field theory. 
We find that SOC does not affect the correlation-induced renormalizations, 
which validates the Hund's metal picture of ruthenates even in the 
presence of the sizeable SOC relevant to these materials.  
Nonetheless, SOC is found to change significantly the electronic structure 
at k-points where a degeneracy applies in its absence. 
We explain why these two observations are consistent with one another
and calculate the effects of SOC on the correlated electronic structure. 
The magnitude of these effects is found to depend on the energy of the quasiparticle state under consideration, 
leading us to introduce the notion of an energy-dependent quasiparticle spin-orbit coupling 
$\lambda^*(\omega)$. This notion is generally applicable to all materials in 
which both the spin-orbit coupling and electronic correlations are sizeable.
%
\end{abstract}
\maketitle

The effect of spin-orbit coupling (SOC) on the electronic properties
of materials is a topic of broad current interest. In weakly
correlated metals, SOC influences band-structure topology, 
and plays a key role for topological insulators and Weyl 
metals~\cite{hasan2010colloquium,wan2011topological}. 
In Mott insulators, SOC influences the atomic multiplet configurations and
the magnetic exchange, which leads to rich spin-orbital
physics~\cite{yosida_book,khomskii_book,jackeli2009mott,khaliullin2013excitonic}.
SOC is also crucial for heavy-fermion compounds, where it selects the multiplet
structure of the localized $f$-electrons~\cite{stewart_rmp,dzero2016topological}. 
In contrast to these limiting cases, the effects of SOC in intermediate 
to strongly correlated metals have been less explored and are less understood.

A notable example~\cite{anwar2016direct,nelson2004odd,kim2017anisotropy,mackenzie2017even,burganov2016strain,hicks2014strong,steppke2017strong} of such a metal is Sr$_2$RuO$_4$, 
which behaves as a strongly correlated Fermi liquid 
below $T_{\mathrm{FL}}\sim 25$~K, 
and undergoes a transition to an unconventional superconducting state below
$1.5$~K~\cite{mackenzie2003superconductivity}.
%
%
The role played by SOC in ruthenates raises an interesting puzzle. 
On the one hand, ruthenates have been successfully described~\cite{mravlje2011coherence} 
as `Hund's metals', a family of compounds in which the electronic correlations 
are driven by Hund's coupling~\cite{georges_Hund_review_annrev_2013}. 
Qualitative understanding and, importantly, quantitatively accurate
results have been 
obtained~\cite{mravlje2011coherence,dang2015band,dang2015electronic,Stricker2016waterfall,
jernej2016thermo} using dynamical mean-field theory
(DMFT~\cite{DMFT_rmp_1996}).
%
While these calculations did not take SOC into account, several 
physical properties were described satisfactorily including 
the measurements of the Knight shift and
$1/T_1$ from NMR, the frequency dependence of the self-energy~\cite{mravlje2011coherence}, 
the optical conductivity~\cite{Stricker2016waterfall} and the Seebeck
coefficient~\cite{jernej2016thermo}.  
In a broader context, direct evidence for the importance of Hund's coupling 
for the physics of ruthenates was recently obtained from photoemission experiments on 
insulating Ca$_2$RuO$_4$~\cite{sutter2016hallmarks}. 
One of the hallmarks of Hund's metals is a large degree of differentiation 
between the different orbitals (here forming the partially occupied 
$t_{2g}$ subshell). Indeed, a much larger mass enhancement is  
observed for $d_{xy}$ than for $d_{xz,yz}$ in Sr$_2$RuO$_4$~\cite{mackenzie03,mravlje2011coherence}. 
Because SOC promotes orbital mixing, such a differentiation would not be observed 
if SOC would dominate the physics. 
Another hallmark of Hund's metals~\cite{yin12,aron15,stadler2015dynamical,horvat2016lowe} 
is that the temperature below which 
the orbital angular momentum no longer fluctuates is much higher 
($\gtrsim 1000$~K in \SRO \cite{jernej2016thermo}) than the 
corresponding temperature for spin degrees of freedom ($\sim$~\TFL), 
pointing at a separation between spin and orbital degrees of freedom 
above \TFL.
All these remarks suggest that SOC might not crucially influence electronic 
correlations in \SRO. 
%

On the other hand, the magnitude of the SOC $\lambda\simeq 0.1$~eV is much larger than 
low-energy scales such as \TFL and is not negligible in comparison to the Hund's rule 
coupling $J_H\approx0.4$~eV.  
Indeed, there is ample theoretical and experimental evidence that SOC 
 does play an important role in Sr$_2$RuO$_4$. 
Band-structure calculations have demonstrated that it influences 
the shape of the Fermi surface (FS) in an important manner~\cite{haverkort2008strong}.  
Angular-resolved photoemission spectroscopy (ARPES) has revealed 
orbital mixing~\cite{iwasawa2010interplay} and sizeable
 lifting of degeneracies~\cite{veenstra2014spin} between bands. 
Recent DMFT calculations at the model level~\cite{kim2016j} and for 
\SRO~\cite{zhang2016fermi} emphasized the importance of SOC, the latter 
work pointing at its enhancement by correlations. 
%

The purpose of the present article is to resolve this apparent conflict. 
%
%
We analyse the renormalizations associated with the orbital-diagonal and 
orbital off-diagonal components of the self-energy and find striking 
differences between them. The former are frequency-dependent (`dynamical') 
but essentially unaffected by SOC. In contrast, the latter 
are static (frequency-independent) over a  wide energy range, corresponding 
to a renormalization of the bare SOC by local interactions. 
This observation provides the solution to the above puzzle. The overall 
dynamical correlations, and in particular the correlation-induced
orbitally-dependent renormalizations remain characteristic of Hund's
metals, while the effect of the SOC is sizeable only 
close to quasi-momenta in the Brillouin zone where degeneracies 
(or quasi-degeneracies) are found for $\lambda=0$.
The calculated electronic structure agrees well with photoemission 
experiments~\cite{iwasawa2010interplay,veenstra2014spin}.
We also provide analytical understanding of the effects of SOC close
to degeneracy points. 
Importantly, because  of the energy dependence of the many-body renormalizations, 
the splitting induced by SOC is found to depend on the energy of the states under consideration.  
This leads us to introduce the concept of an {\it energy-dependent quasiparticle spin-orbit coupling}, 
$\lambda^*(\omega)$.


Previous work on Sr$_2$RuO$_4$ disregarded the effects of SOC, with the exception of 
Ref.~\cite{zhang2016fermi} who found that SOC is enhanced by 
the off-diagonal components of the self-energy.  
However, the authors of Ref.~\cite{zhang2016fermi} did not consider the properties of states 
away from the Fermi level and hence did not reveal the energy-dependence of the effect of SOC 
on quasiparticles, as described by $\lambda^*(\omega)$. This is crucial to understand the 
effect of SOC on the ARPES spectra of ruthenates~\cite{veenstra2014spin} but also for other materials, 
such as iron-based superconductors~\cite{borisenko16,day2017ubiquitous}.


\begin{table}[t]
\caption{Hopping amplitudes $t_{ab}^{{\textbf{R}}}$ between orbitals
  ($a,b=xy,xz,yz$) along a lattice vector $\textbf{R}$, entering the
  tight-binding hamiltonian $H_0$.  We use $t$=0.42, $t_{1}$=0.17,
  $t_{2}$=0.30, $t_{3}$=0.03, and $t_{4}$=0.04 (eV)~\cite{okamoto2004electron}, and crystal field
  term $\epsilon_{\mathrm{CF}} (n_{xz}+n_{yz})$ with
    $\epsilon_\mathrm{CF}=0.1$ eV. Due to the Hund's rule coupling, the
    results are robust with respect to the changes of
    crystal-field.~\cite{Note1}}
\begin{ruledtabular}
\begin{tabular}{l  c  c  c  c  c}
           ~ & ~$xz,xz$
             & ~$yz,yz$
             & ~$xy,xy$
             & ~$xz,yz$~\\
$t_{ab}^{\pm1,0,0}$    & $-t_{2}$ & $-t_{3}$ & $-t$ & 0\\
$t_{ab}^{0,\pm1,0}$    & $-t_{3}$ & $-t_{2}$ & $-t$ & 0\\
$t_{ab}^{\pm1,\pm1,0}$       & 0 & 0 & $-t_{1}$ & $-t_{4}$\\
$t_{ab}^{\pm1,\mp1,0}$      & 0 & 0 & $-t_{1}$ & $t_{4}$\\
\end{tabular}
\label{table:hopping_term}
\end{ruledtabular}
\end{table}
\begin{figure}[b]
\includegraphics[width=\columnwidth]{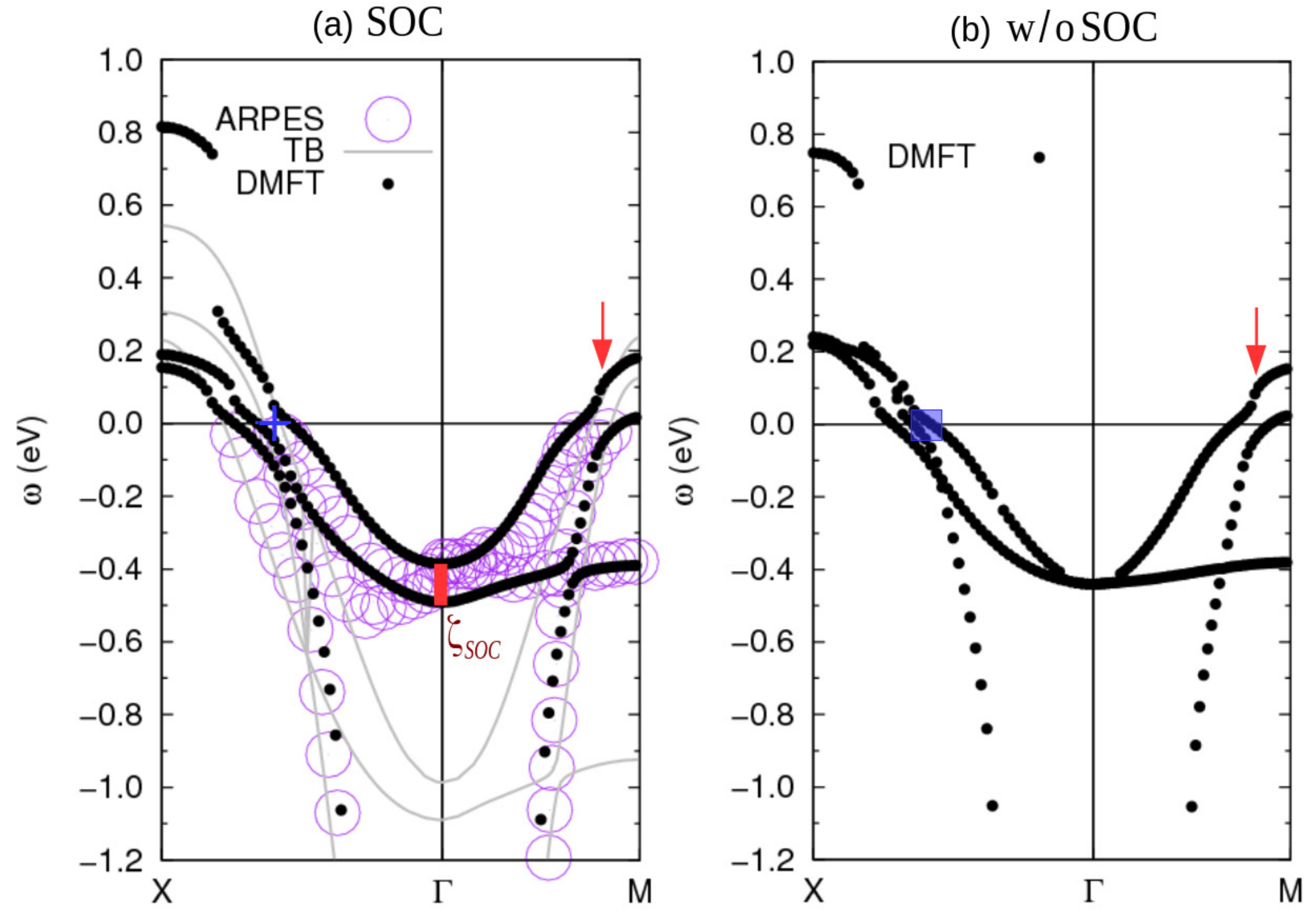}
\caption{ 
(a) DMFT+SOC quasiparticle dispersions along the X-$\Gamma$-M path (black dots). 
The ARPES measurements~\cite{iwasawa2010interplay} are displayed as open circles, and   
the bands from TB+SOC in the absence of interactions as light gray lines.   
The SOC-induced splitting at the $\Gamma$ point is indicated as a vertical bar. 
The cross at Fermi level along the $X$-$\Gamma$ path indicates the SOC-induced 
band splitting near the Fermi level. 
(b) Same as (a) but for DMFT without SOC. 
The square indicates the point where the $\beta$ and $\gamma$ sheets are nearly degenerate.   
The arrows indicate Fermi surface crossings along $\Gamma$-M at which the effect of SOC is small.
\label{fig:ARPES1}
}
\end{figure}

We describe Sr$_2$RuO$_4$ using the model Hamiltonian, $H=H_{\rm 0}+H_{\rm ls}+ H_{\rm int}$. 
$H_{\rm 0}$ is a TB hamiltonian
with parameters specified in Table~\ref{table:hopping_term}, see also SM~\footnote{See the Supplemental Material for (i) the
      insensitivity of the crystal-field independent electronic
      structures, (ii) discussion of our implementation of the solver,
      (iii) SOC effects on the momentum distribution curve at $\omega$=0, (iv) analytical discussion of SOC induced splitting
      of quasiparticle bands, (v) Matsubara self-energies and the
      discussion of the temperature dependence, and (vi) plots of FS
      for cases without SOC and without correlations.}.  
In the basis having $m$ as a quantum number of $l_{z}$ ($l_{z}$ is 
the angular momentum operator projected on $t_{2g}$ manifold), the SOC is described by:
\begin{eqnarray}
H_\mathrm{ls}=&& \frac{\lambda_z}{2}\sum_{m=-1}^1
m(d_{m\uparrow}^\dagger d_{m\uparrow} - d_{m\downarrow}^\dagger
d_{m\downarrow}) \label{eq:SOC}\\ &+&\frac{\lambda_{xy}}{\sqrt{2}}\sum_{m=-1}^{0}
(d_{m+1\downarrow}^\dagger d_{m\uparrow} + d_{m\uparrow}^\dagger
d_{m+1\downarrow})\nonumber
\end{eqnarray}
with
$\lambda_\mathrm{xy}=\lambda_\mathrm{z}=\lambda=100$\,meV. (Distinct
symbols $\lambda_\mathrm{xy},\lambda_\mathrm{z}$ are introduced for
later convenience). 
The Coulomb interactions $H_{\rm int}$ within the $t_{2g}$
manifold are described using the rotationally
invariant Kanamori Hamiltonian $H_{\rm int}=(U-3\J_{\rm
  H})\hat{N}(\hat{N}-1)/2-2\J_{\rm H} \vec{S}^{2}-{\J_{\rm H}} \vec{L}^2/2$ with
$\hat{N}$,$\vec{S}$,$\vec{L}$ being the atomic charge, spin and angular momentum
operators. 
We used $U$=2.3 eV and $J_{H}$=0.4 eV, 
as established in previous work~\cite{mravlje2011coherence} 
and fixed the occupancy to $\langle \hat{N}\rangle =4$, as relevant to ruthenates.
$H$ was solved in the framework of DMFT, using the hybridization
expansion continuous-time quantum Monte Carlo (CTQMC)
solver~\cite{gull_CTQMC_rmp_2011}, as implemented in the TRIQS
library~\cite{TRIQS_Package_2015,TRIQS_CTQMC_2016}, with 
complex-valued imaginary time Green's functions, as required in general when handling SOC~\cite{Note1}.
The analytical continuation of Matsubara self
energies to the real axis was performed using the 
stochastic analytical continuation~\cite{beach2004identifying}. The temperature of simulations
was 230\,K for all the results described in the main text.

We first discuss the basic phenomenology emerging from the calculated electronic structure, 
as displayed on Fig.~\ref{fig:ARPES1}(a).  
The filled symbols indicate the location of the maxima of the spectral functions, 
obtained from DMFT including SOC, 
along the two directions $\Gamma$-$X$ and $\Gamma$-$M$ in the Brillouin zone.   
For comparison, the non-interacting band-structure from the TB$+$SOC Hamiltonian 
is shown with thin lines, as well as the ARPES data by Iwasawa et al~\cite{iwasawa2010interplay}.  
The basic experimental observations are well reproduced by DMFT. 
We note in particular:  
(i) the strong quasiparticle mass renormalizations near Fermi level and 
their marked orbital dependence~\cite{burganov2016strain,mravlje2011coherence}: $\sim 4.0$ for $xy$, $\sim 3.1$ 
for $xz,yz$~\footnote{The renormalizations depend on temperature and increases towards 
the measured ones when the temperature of the simulation is reduced.~\cite{deng13}}.
(ii) `Unrenormalization' of the quasiparticle dispersion as the binding energy 
is increased, with effective velocities becoming closer to the bare ones.  
This effect is accompanied by waterfall structures~\cite{iwasawa2012high} - an effect which 
has been observed in measurements of the optical conductivity~\cite{Stricker2016waterfall} 
and also visible on Fig.~\ref{fig:ARPES1} close to X point.
(iii) Because of these unrenormalizations, the overall bandwidth of the $xz/yz$ quasiparticle 
band is about half of the non-interacting one, whereas that of the $xy$-band is
approximately equal to the non-interacting one (despite the strong
low-energy mass renormalizations). 
These unrenormalizations will
  play an important role also for the energy dependent effects of SOC 
  $\lambda^*(\omega)$. 
  

\begin{figure}[t]
\includegraphics[width=.8\columnwidth]{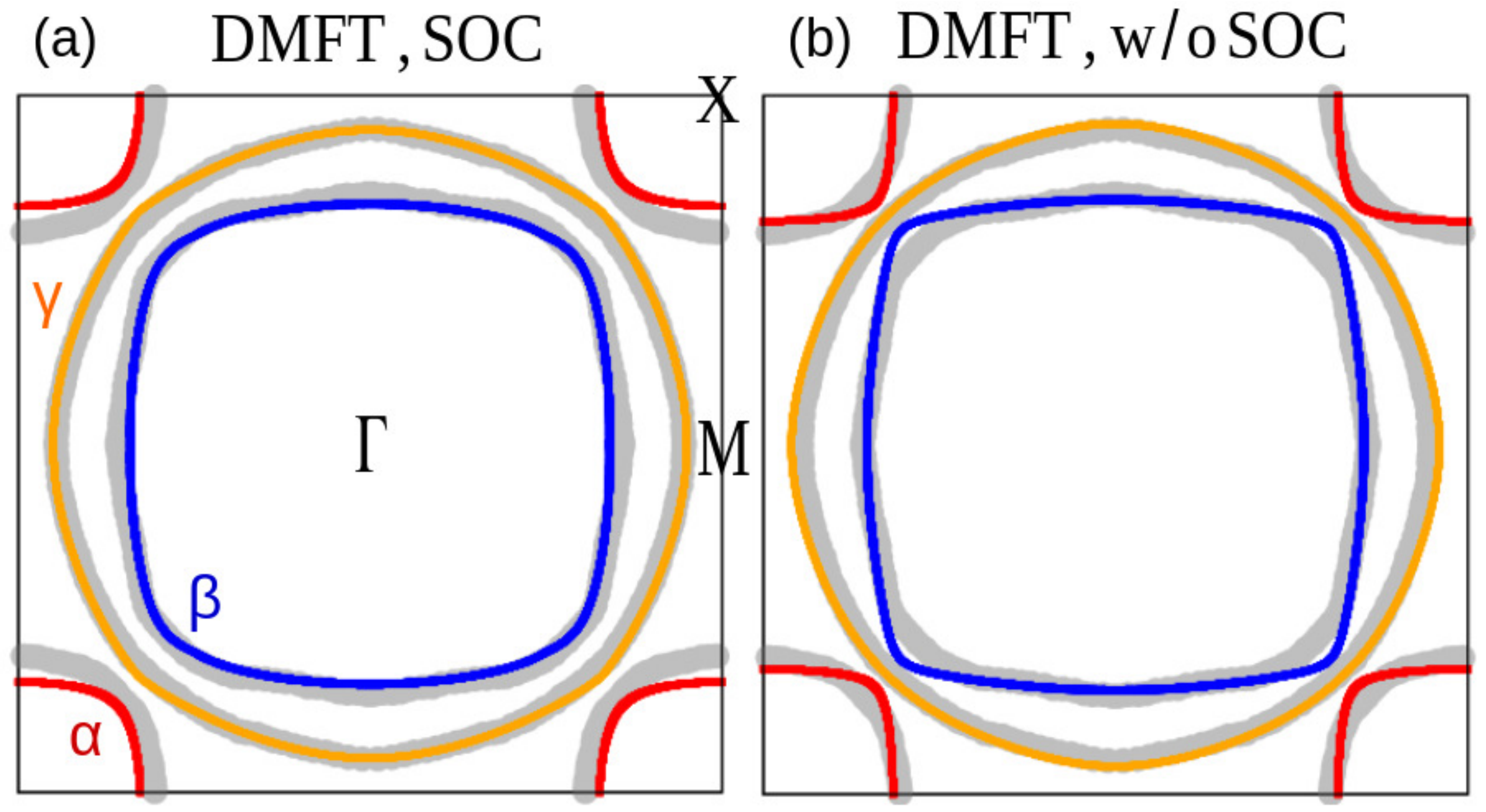}
\caption{ FSs of Sr$_{2}$RuO$_{4}$ in the DMFT with included SOC (a) and
DMFT without SOC (b).
The blue, orange, and red lines present
$\beta$, $\gamma$, and $\alpha$ sheets.
The FS from the ARPES experiment Ref.\cite{damascelli2000fermi}
is also reproduced (gray).
\label{fig:FS} }
\end{figure}

A comparison with Fig.~\ref{fig:ARPES1}(b) 
(displaying our DMFT results without SOC, consistent with previous work~\cite{mravlje2011coherence,Stricker2016waterfall}) 
reveals that all the basic features (i-iii) are already correctly described without taking SOC into account.  
Nonetheless, the SOC has several important consequences that we now discuss. 
As apparent from Fig.~\ref{fig:ARPES1}(a,b), the effect of SOC is strongest 
at (near-)degeneracy points. We emphasize in particular: 
(i) at the $\Gamma$-point and for a binding energy $\sim -0.5$~eV, the lifting of degeneracy 
by $\zeta_\mathrm{SOC}\simeq 106$~meV between the two bands originating from the $xz,yz$ orbitals 
(signalled by a red bar on Fig.~\ref{fig:ARPES1}a),  
and 
(ii) along the $\Gamma$-$X$ direction, at the Fermi level, the $\beta$ and $\gamma$ sheets 
of the Fermi surface almost touch when SOC is neglected (as indicated by the blue square 
on panel (b) of Fig.~\ref{fig:ARPES1}). The inclusion of SOC lifts this degeneracy 
by an amount $\zeta'_\mathrm{SOC}\simeq$\,104 meV (blue cross in panel (a)). Correspondingly, the 
Fermi momenta are shifted by an amount $\pm \delta_k$ and, as shown below, the 
Fermi velocities of the two sheets become equal. 
These effects are typical of level repulsion. 
Away from the near-degeneracies the effect of SOC is smaller: 
the dispersions along $\Gamma$-$M$ and the
corresponding Fermi surface crossings (around the red arrows on
Fig.\ref{fig:ARPES1}) are affected less. In particular, the Fermi
velocities and the magnitude of the renormalizations are orbital/sheet-dependent.

The calculated FS including SOC is displayed on Fig.~\ref{fig:FS}(a) and compared to the measured one~\footnote{
Longer range hoppings and out-of-layer tunneling terms would improve quantitative agreement with experiment.}. 
The  SOC modifies the Fermi surface, leading to an inflation of the $\gamma$-sheet.  
This inflation is not due to a change in 
orbital populations, which remain basically unchanged from their $\lambda=0$ values 
$0.64$ and $0.68$ (per spin) for $xy$ and $xz$, respectively.
Instead, it is driven by the admixture of $xz/yz$ orbital character into the $\gamma$ sheet.
The fact that orbital polarization in a DMFT calculation with SOC remains similar to the one
found without taking SOC into account is another demonstration that
the Hund's rule coupling physics is not affected by SOC.~\cite{Note1} 

\begin{figure}[t]
\includegraphics[width=\columnwidth]{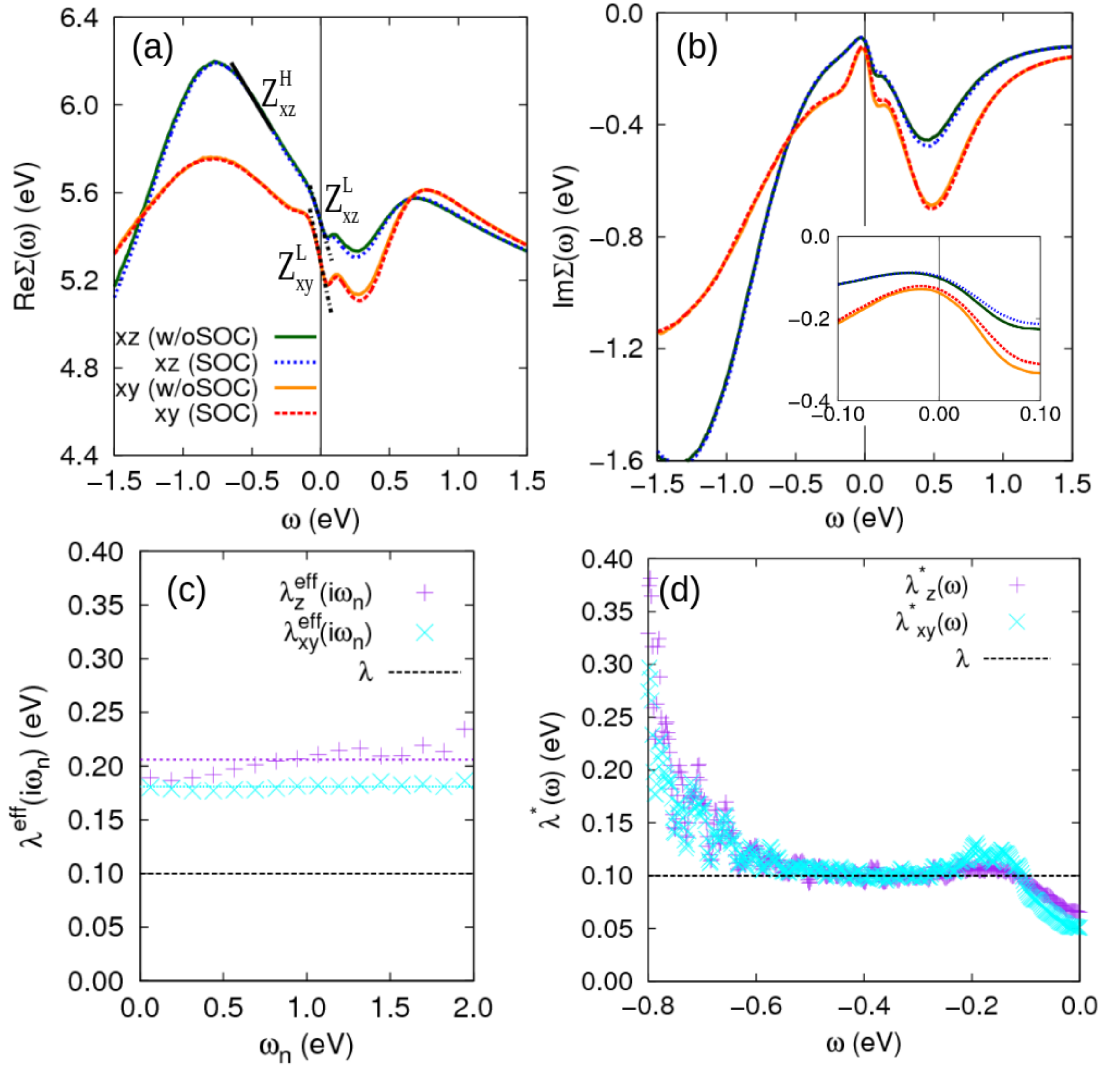}
\caption{
Real (a) and imaginary (b) parts of the real-frequency 
orbital-diagonal components of the self-energy (with SOC dashed, without SOC plain).
Two linear fits $\mathrm{const} + \omega(1-1/Z)$ are also displayed in (a), corresponding to low-frequency 
(dashed line, slope set by $Z^{\mathrm{L}}_{a}$) and higher frequency 
around $\omega=-0.5$~eV (plain line, slope set by $Z^{\mathrm{H}}_{a}$). 
(c): Off-diagonal components of the Matsubara self-energy, presented as an effective renormalization of SOC 
(see text). Note the weak frequency dependence. The straight horizontal lines denote 
average values over the range $[0,2]$~eV in comparison to the bare value 
$\lambda=0.1$~eV.  
(d) Energy dependent quasiparticle SOC 
$\lambda^*_{z} (\omega)$
and $\lambda^*_{xy} (\omega)$ 
(see text).
\label{fig:Self_energies}
}
\end{figure}

In order to understand why the overall effect of correlations are
unchanged by SOC while the fermiology close to degeneracy points is
affected significantly, we examine the DMFT self-energies. The
orbital-diagonal components are displayed in panels (a,b) of
Fig.~\ref{fig:Self_energies}, in the presence and absence of SOC,
respectively.  Strikingly, there is very little difference between the
two calculations.  All the key features
(the linear low-frequency behaviour of the real parts $\propto
(1-1/Z_{a})\omega$, followed by a kink at $\sim-50$ meV, the features
at positive energies, the frequency dependence of the imaginary
part/scattering rate) are not only found to be similar but practically
the same.  Such insensitivity of electronic correlations to the SOC
can be rationalized in terms of the high energy scale below which
orbital fluctuations are screened, characteristic of Hund's metals~\cite{jernej2016thermo,horvat2017spin}.

The effect of SOC on electronic correlations becomes apparent
only when looking at the off-diagonal components of the self-energies 
$\Sigma_{xz;xy}$ and $\Sigma_{xz;yz}$. 
These vanish by symmetry for $\lambda=0$ but acquire a finite value once
SOC is included. 
On Fig.~\ref{fig:Self_energies}(c) we display, as a function of Matsubara frequency $\omega_n$, 
the quantities~\footnote{Equivalently, $\lambda^{\mathrm{eff}}_{xy}$ can be defined from 
$\mathrm{Re}\Sigma_{yz\uparrow,xy\downarrow}$, which is identical 
to $\mathrm{Im}\Sigma_{xz\uparrow,xy\downarrow}$ by symmetry.  
}:   
$\lambda^{\mathrm{eff}}_{z}(i\omega_{n})=\lambda_{\mathrm{SOC}}+2\mathrm{Im}\Sigma_{xz\uparrow,yz\uparrow}(i\omega_{n}),
\lambda^{\mathrm{eff}}_{xy}(i\omega_{n})=\lambda_{\mathrm{SOC}}-2\mathrm{Im}\Sigma_{xz\uparrow,xy\downarrow}(i\omega_{n})$. 
As seen there, the off-diagonal components are frequency-independent over a 
wide frequency range (exceeding $\sim 2$~eV), hence they can be
considered as  additional single-particle terms that add-up to the bare ones~\cite{zhang2016fermi}, yielding $\lambda^{\mathrm{eff}}_{z}=206$~meV, $\lambda^{\mathrm{eff}}_{xy}=181$~meV, enhanced by about a factor of two over the bare value $\lambda=0.1$~eV.

We are now in a position to perform a more quantitative analysis of the effects of SOC.
Let us focus first on quasiparticle dispersions, which are 
the solutions $\omega(k)$ of:  
\begin{equation}\mathrm{det}[(\omega+\mu)\delta_{\abar,\bbar}-H^0(k)_{\abar,\bbar}-
\mathrm{Re} \Sigma(\omega)_{a} \delta_{\abar,\bbar}- \hat{\lambda}^{\mathrm{eff}}_{\abar,\bbar}]=0 
\label{eq:qp}
\end{equation} 
in which here $\abar,\bbar$ are indices for the six possible spin-orbital combinations $(a\sigma)$. 
Close to the Fermi level, this equation can be simplified by inserting the linear frequency 
dependence of the self-energy 
$\mathrm{Re}\Sigma_a=\Sigma_a(0)+\omega(1-1/Z_{a})+\cdots$ and linearizing in the momentum-dependence 
close to the FS. 
Considering for example the $\Gamma$-$X$ direction for which the $\beta$ and $\gamma$ bands are almost 
degenerate close to the FS in the absence of SOC (Fig~\ref{fig:ARPES1}), and considering the $\alpha$-band to be lower in energy, 
an analytical solution to Eq.~(\ref{eq:qp}) can be obtained~\cite{Note1}. 
This leads to the following expression for the SOC-induced splitting in momentum $\pm \delta_k$ 
and energy $\zeta^\prime_{\mathrm{SOC}}$, 
and for the renormalized quasiparticle velocities:
\begin{eqnarray} \label{eq:zetaprime}
&\delta_k = \frac{\leff_{xy}}{2\sqrt{v_{\beta }v_{\gamma }}}=\frac{\sqrt{Z_{xz} Z_{xy}}\leff_{xy}}{2\sqrt{v^*_{\beta }v^*_{\gamma }}}\,\,,\,\,
v^* = 2\frac{v^*_{\beta}v^*_{\gamma}}{v^*_{\beta}+v^*_{\gamma}}\\
&\zeta^\prime_{\mathrm{SOC}}=2 v^*\delta_k = 2 \frac{\sqrt{v^*_{\beta}v^*_{\gamma}}}{v^*_{\beta}+v^*_{\gamma }}
\sqrt{Z_{xy}Z_{xz}}\,\leff_{xy} 
\,=\,\frac{v^*}{\sqrt{v^*_{\beta}v^*_{\gamma}}} \lambda^*_{xy}(0) \nonumber 
\end{eqnarray}

\begin{figure}[h]
\includegraphics[width=\columnwidth]{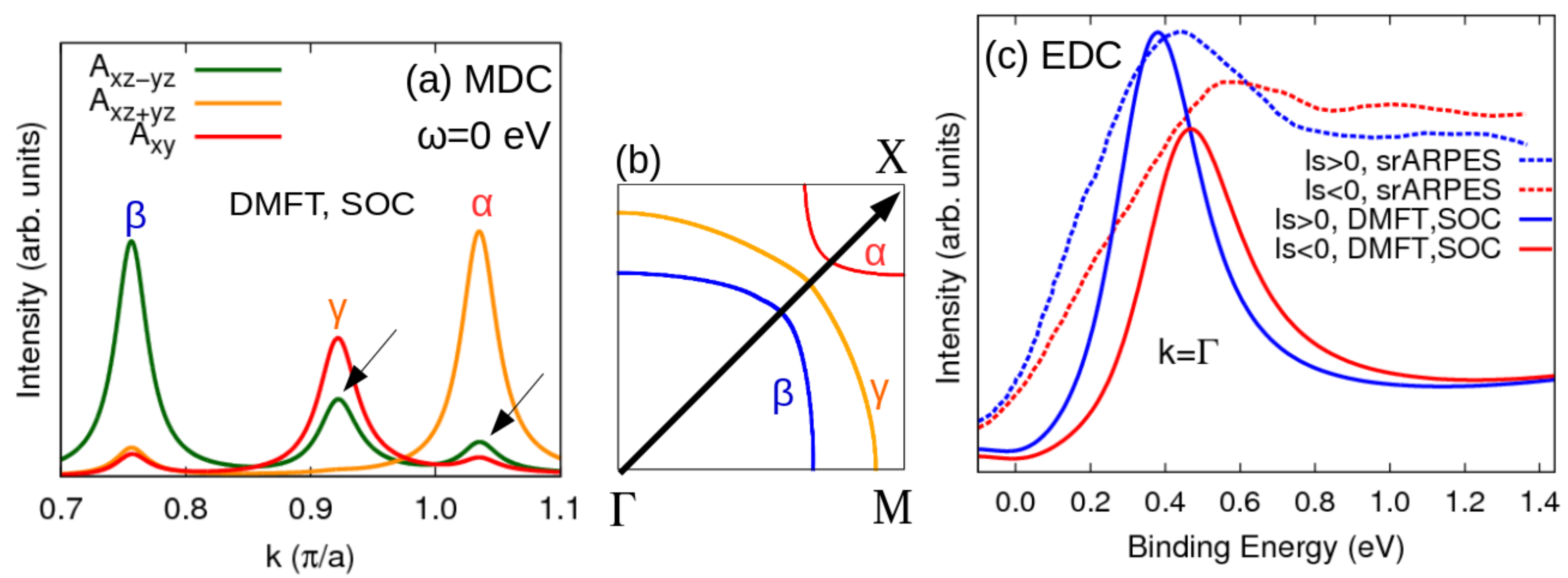}
\caption{
(a) Momentum distribution curves (MDC) at zero energy $\omega$=0 along
the $\Gamma$-X path, projected on the $xy$, $xz-yz$ and $xz+yz$ orbital components. 
The arrows emphasize the non-zero contribution of 
the $xz-yz$ orbital to the $\gamma$ and $\alpha$ bands.  
(c) Energy distribution curves (EDC) at the $\Gamma$-point, for 
states with spin parallel ($ls>0$) and anti-parallel ($ls<0$)
to the orbital moment, compared to spin-resolved ARPES results~\cite{veenstra2014spin}. 
A Gaussian broadening of $0.05$~eV was used.
\label{fig:ARPES2}
}
\end{figure}

In this expression, $v_{\beta},v_{\gamma}$ are the bare TB Fermi
velocities in the absence of SOC and interactions for the $\beta$ and
$\gamma$ sheets, and $v^*_{\beta}=Z_{xz} v_{\beta},v^*_{\gamma}=Z_{xy}
v_{\gamma}$ are the renormalized quasiparticle velocities in the
presence of interactions but without SOC.  
We have introduced energy-dependent quasiparticle spin-orbit couplings, defined by:
\begin{equation}
\lambda^*_{z} (\omega)= Z_{xz}(\omega) \lambda^{\mathrm{eff}}_{z}
\,\,\,,\,\,\, 
\lambda^*_{xy}=\sqrt{Z_{xy}(\omega) Z_{xz} (\omega)}\lambda^{\mathrm{ eff}}_{xy}
\end{equation}
The lifting of degeneracy by SOC equalizes the velocities for each sheet, hence suppressing the
orbital/sheet differentiation at this specific k-point, an effect
clearly visible on Fig.~\ref{fig:ARPES1}(a), which deserves to be resolved from high resolution ARPES experiments. 
We find $\delta_k \simeq
0.087 \pi/a$ and $\zeta^\prime_{\mathrm{SOC}}\simeq 104$~meV.  Away
from degeneracy points (e.g. for FS crossings along $\Gamma$-M), the
SOC can be treated perturbatively and only modifies weakly the Fermi
momenta and quasiparticle velocities by an amount of order
$(\leff)^2/\Delta\varepsilon$, with $\Delta\varepsilon$ the
quasiparticle energy separation between the considered band and the
one closest to it in energy. 
Note also that the effective coupling  $\lambda^*_{xy}(0)$ which 
enters $\zeta^\prime_{\mathrm{SOC}}$ is suppressed by the 
renormalization factor $\sqrt{Z_{xy} Z_{xz}} \approx 1/4$ as compared to $\lambda^\mathrm{eff}$. 
The electronic correlations thus, on one hand, increase the magnitude of the SOC through the orbital off-diagonal
self-energy but, on the other hand, suppress it due to the quasiparticle renormalization factor.  

Further insight into the orbital content of each FS sheet along the 
$\Gamma$-X direction can be obtained by looking at 
the momentum-distribution curves (MDCs), which are displayed on 
Fig.~\ref{fig:ARPES2}(a) by using a projection onto the $xy$, 
$xz-yz$ and $xz+yz$ orbital components. 
We observe that the $\gamma$-sheet crossing has almost equal 
contributions from $xy$ and $xz-yz$ orbital components, 
as previously discussed from electronic structure calculations~\cite{haverkort2008strong} 
and ARPES experiments~\cite{iwasawa2010interplay}.  
The fact that all sheets have a non-vanishing $xz-yz$ component  
(odd under mirror-plane $x\leftrightarrow y$ symmetry) is consistent 
with the ARPES results of Ref.~\cite{iwasawa2010interplay} for 
sigma-polarized light. 

The quasiparticle equation (\ref{eq:qp}) can also be used to discuss the SOC-induced 
$\Gamma$-point splitting between the $xz$ and $yz$-dominated bands at a binding energy $\omega_\Gamma \sim -0.5$~eV. 
The key difference here 
(in contrast to the energy splitting at the FS, $\zeta^\prime_{\mathrm{SOC}}$) 
is that the quasiparticle renormalization appropriate to this 
higher energy must be used (Fig.~\ref{fig:Self_energies}), which is given by 
$\left[(1-d \mathrm{Re} \Sigma_{xz}/ d \omega)|_{\omega=\omega_\Gamma}\right]^{-1}=Z_{xz}(\omega_\Gamma) = 0.49$, 
different (and larger) from the low-energy values $Z_{xz}=Z_{yz}\simeq 0.32$. 
As a result, the SOC-induced $\Gamma$-point splitting reads 
\begin{equation}
\zeta_{\mathrm{SOC}}= Z_{xz}(\omega_\Gamma) \lambda_{z}^{\mathrm{eff}} =  \lambda_z^*(\omega_\Gamma).
\label{eq:Gamma}
\end{equation}
The energy dependence of the quasiparticle renormalization $Z(\omega)$
thus causes the effects of the SOC also to be energy dependent, and given by $\lambda^{*}(\omega)$. 
This energy-dependent SOC acting on quasiparticles states is plotted on Fig.~\ref{fig:Self_energies}(d).

We note that at $\omega_\Gamma$ there is a compensation between the
correlation-induced enhancement $\leff_z\simeq 2\lambda$ and the
renormalization by $Z^\mathrm{H}_{xz}\simeq 1/2$, so that we obtain
$\zeta_{\mathrm{SOC}}\simeq 106$~meV to be quite close to the bare
$\lambda$. It is only due to this accidental cancellation that the LDA
\cite{haverkort2008strong} correctly predicted the observations of
spin-resolved ARPES~\cite{veenstra2014spin}. Last, we turn to a closer
comparison of our result to this experiment that selectively probed
electrons with orbital angular momentum parallel ($ls>0$) and
anti-parallel ($ls<0$) to the electron spin~\cite{veenstra2014spin}.
Hence, we display in Fig.~\ref{fig:ARPES2}(c) the calculated
$ls$-resolved spectral functions at the $\Gamma$ point. The location
of the maxima of the two peaks, their splitting
$\zeta_{\mathrm{SOC}}\simeq 106$~meV and the lower intensity of the
higher binding energy $ls<0$ peak are well reproduced.

 In summary, there are three qualitative take-home messages from our work. 
 First, because the SOC is smaller than the orbital coherence scale, 
 it does not affect the dynamical properties of \SRO, and electronic correlations 
 are characteristic of a Hund's metal. 
 This validates previous theoretical work that neglected SOC~\cite{mravlje2011coherence,kim2015nature,Stricker2016waterfall,dang2015band,dang2015electronic,
 jernej2016thermo,han2016ferromagnetism,dasari2016first,deng2016transport}. 
 Second, the SOC does change significantly the electronic structure 
 at k-points where a degeneracy is found in the absence of SOC. 
 Thirdly, when evaluating these effects, it is crucial to take into account 
 both the static renormalization of the SOC by correlations, 
 and the quasiparticle renormalization factors taken in the appropriate energy range,  
 leading to the notion of an energy-dependent quasiparticle SOC. 
 These observations are  important for the ongoing discussion of the superconducting order
 parameter of \SRO where SOC plays a key role~\cite{akbari2016SF,kim2017anisotropy}, 
 and are of general relevance to a broad class of materials involving SOC and electronic correlations.

\acknowledgements
We are grateful to M.~Aichhorn, F.~Baumberger, G.~Kraberger and A.~Tamai for useful discussions 
and to the CPHT computer support team. 
This work was supported by the European Research Council grants ERC-319286-QMAC (M.K., A.G.) 
and ERC-278472-MottMetals (O.P.), and by the Swiss National Science Foundation (NCCR MARVEL). 
The Flatiron Institute is supported by the Simons Foundation. 
J.M. acknowledges support by the Slovenian Research Agency (ARRS) under Program P1-0044.
\bibliography{refs_Sr2RuO4_SOC.bib}

\end{document}